\documentclass[twocolumn,aps,floatfix,superscriptaddress]{revtex4-2}
\usepackage[utf8]{inputenc}

\usepackage[inline]{enumitem}
\usepackage{graphicx,color}
\usepackage{dcolumn}
\usepackage{epsfig}
\usepackage{euscript}
\usepackage{amssymb}
\usepackage{amsmath}
\usepackage{amsthm}
\usepackage{newlfont}
\usepackage{bm,mathrsfs}
\usepackage{subfigure}
\usepackage{xcolor}
\usepackage{changes}
\setaddedmarkup{\textcolor{red}{#1}}
\usepackage{todonotes}
\usepackage{subfigure}
\usetikzlibrary{calc}
\usepackage[colorlinks,linkcolor=blue,urlcolor=blue,citecolor=blue]{hyperref}
\usetikzlibrary{arrows.meta}
\newcommand{\opunit}{\textrm{1}\kern-0.22em\textrm{l}}

\usepackage{makeidx}
\usepackage{comment}

\def\bea{\begin{eqnarray}}
\def\eea{\end{eqnarray}}
\def\ba{\begin{array}}
\def\ea{\end{array}}

\def\bea{\begin{eqnarray}}
\def\eea{\end{eqnarray}}
\def\ba{\begin{array}}
\def\ea{\end{array}}

\def\la{\langle}
\def\ra{\rangle}

\definecolor{dgreen}{rgb}{0,0.7,0}

\begin{document}

\title{Resetting in a viscoelastic bath: the bath remembers}
 \author{Ion Santra}
 \affiliation{Department of Physics and Astronomy, KU Leuven }
 \author{Debankur Das}
  \affiliation{Institut f\"ur Theoretische Physik, Universit\"at G\"ottingen }

\begin{abstract}
We study stochastic resetting of a probe particle in a viscoelastic environment where only the probe is reset while the medium retains memory of its past dynamics. Using a minimal model with finite correlation time, we analyze the competition between the resetting timescale and the viscoelastic relaxation timescale. This interplay leads to nonequilibrium steady states that differ qualitatively from those of Markovian Brownian motion with resetting. In particular, strong memory effects produce stationary position distributions with non-exponential tails. For instantaneous resets, we derive the limiting steady-state distributions analytically and compute exactly the time dependent leading non-vanishing moments. We also investigate non-instantaneous resetting via constant-velocity return protocols. In contrast to overdamped Brownian motion, where steady-state fluctuations are independent of the return dynamics, we find that in a viscoelastic medium the fluctuations depend on the reset velocity. This protocol dependence arises from the finite memory of the environment and highlights the role of environmental correlations in resetting-induced steady states. 
\end{abstract}

\maketitle
\section{Introduction}
\label{sec:intro}
Stochastic resetting, where a dynamical process is intermittently stopped, and restarted from a prescribed configuration or distribution, has emerged as a simple yet powerful mechanism for controlling stochastic dynamics~\cite{evans2020stochastic,gupta2022stochastic}. 
For a Brownian particle, resetting suppresses the diffusive spreading and generates non-Gaussian stationary states with exponential tails. Moreover, resetting at an optimal rate can minimize the mean first-passage time to an absorbing boundary~\cite{evans2011optimal,reuveni2016optimal,evans2020stochastic,gupta2022stochastic}. These properties have made stochastic resetting a useful framework for modeling and optimizing search processes~\cite{basu2024target,sunil2024minimizing,pal2024random,munoz2025learning,biswas2025target,chechkin2018random,fuchs2016stochastic}. Stochastic resetting has also been explored in a variety of physical systems, including reaction--diffusion processes~\cite{durang2014statistical}, active matter~\cite{evans2018run,kumar2020active,santra2020run,tucci2022first,bressloff2020occupation,pal2024active,scacchi2018mean}, fluctuating interfaces~\cite{gupta2014interfaces}, and interacting many-body systems~\cite{nagar2023stochastic,magoni2020ising,sadekar2020zero,acharya2025manipulating}. Beyond physics, resetting ideas have also been applied to problems in biology~\cite{roldan2016stochastic}, computer algorithms~\cite{luby1993optimal,montanari2002optimizing}, and social sciences~\cite{santra2022effect,bonomo2022mitigating,jolakoski2025impact}. In addition to these applications, a variety of extensions of stochastic resetting have been explored. These include effects on inertial dynamics~\cite{gupta2019underdamped,olsen2024velocity}, processes evolving in external potential landscapes~\cite{pal2015diffusion}, models with fluctuating parameters such as diffusion coefficients or stochastic energy renewal~\cite{santra2022diffusion,santra2025energyrenewal}, resetting to random distributions~\cite{olsen2023steady}, and partial resetting protocols~\cite{tal2022diffusion,di2023time}. Multiparticle simultaneous stochastic resetting has recently emerged as a minimal setting to generate dynamically emergent correlations~\cite{biroli2023extreme,de2026dynamically}, where correlations arise from the dynamics rather than built-in interactions.

Earlier works on stochastic resetting largely considered the limit of instantaneous resets, in which the particle is reset to a fixed position without any time cost. Such a setting enables exact analytical treatments and has revealed a number of universal features of reset-driven dynamics~\cite{smith2023striking}. More recently, attention has shifted toward non-instantaneous resetting, which is often more realistic in experimental implementations~\cite{besga2020optimal,tal2020experimental}. Here the system returns to the reset position through a deterministic or controlled motion e.g. constant velocity or acceleration~\cite{pal2019time,pal2019invariants,bodrova2020resetting,radice2022diffusion,radiceacc}, or via switching trapping potentials~\cite{santra2021brownian,mercado2022reducing}. Experiments with colloidal particles in feedback-controlled optical potentials~\cite{ginot2026experimental,vatash2025many,besga2020optimal} have demonstrated several of these mechanisms stimulating further theoretical work on resetting with explicit return dynamics.~\cite{gupta2021stochastic,maso2019transport}. 

In nearly all of these studies, however, the underlying particle motion is assumed to be Markovian, as in ordinary Brownian diffusion where the medium responds instantaneously to the particle’s motion. Under this assumption, each reset effectively renews the dynamics and the particle’s previous trajectory becomes irrelevant, and successive excursions between reset events are statistically independent.  Models of stochastic resetting with memory have also been explored, where the particle resets to previously visited states rather than to a fixed configuration~\cite{boyer2024active}. Many realistic environments depart from this picture e.g. viscoelastic fluids where the medium can retain memory of the particle’s past motion~\cite{cates1990statics}, producing non-Markovian dynamics that are naturally described by generalized Langevin equations with memory kernels~\cite{caspers2025stochastic,muller2020properties,caspers2023mobility}. Such viscoelastic responses are common in polymeric~\cite{ferry1961viscoelastic} and biological media, including the cytoplasm and other crowded cellular environments~\cite{sato1983rheological}, where slow stress relaxation leads to long-lived correlations~\cite{muller2020properties} in particle motion. These memory effects are known to alter transport properties substantially, giving rise to behaviors such as anomalous relaxation~\cite{gomez2015transient,winter2012active,vaidya2025observation}, shear thinning~\cite{jain2021two,wilson2011microrheology}, and Magnus responses~\cite{cao2023memory} of driven particles. These memory effects should be distinguished from the persistence often discussed in active matter systems, where correlations arise from internal propulsion dynamics, and can be realized in terms of colored noises with different dynamical properties~\cite{santra2022universal}. In viscoelastic media, by contrast, the correlations originate from the delayed mechanical response of the surrounding environment.

As a consequence of quiescent fluid retaining memory, correlations generated before a reset survive and may influence the motion that follows. As a result, their properties differ qualitatively from their Markovian counterparts. Thus, while stochastic resetting in Markovian systems provides an important foundation, understanding its consequences in systems with strong environmental memory is essential for describing such realistic conditions. Initial steps toward addressing this problem have been taken in recent works. In Ref.~\cite{biswas2025resetting}, stochastic resetting was studied for a tracer governed by a generalized Langevin equation, although in that framework the reset operation also restored the bath degrees of freedom, effectively erasing the accumulated memory of he environment between resets. More recently, resetting in a viscoelastic environment was realized experimentally using colloidal particles in optical traps~\cite{ginot2026experimental}, demonstrating that environmental memory can oppose the reset and generate correlations in the particle dynamics. While these studies highlight the importance of memory effects, a systematic understanding of stochastic resetting in genuinely viscoelastic, non-Markovian media remains largely incomplete.

In this work, we address this gap by studying a resetting tracer in a viscoelastic media where we reset the tracer only. Specifically, we consider a probe particle coupled to an auxiliary coordinate representing the viscoelastic medium; the auxiliary degree of freedom evolves continuously and is not affected by resetting. Resetting is applied exclusively to the probe, reflecting experimentally realistic situations in which only the observable colloid is actively manipulated while the surrounding bath remains unperturbed. 
As a result the correlations stored in the bath persist across resetting events, so memory accumulated prior to a reset continues to influence the probe dynamics afterward. We illustrate that this has profound consequences on the dynamical and stationary fluctuations of the tracer position, which become very different from the Markovian case. We first discuss the case of instantaneous resetting by studying the time-dependent position variance analytically which shows a two-step decay. We also study the stationary position distribution of the tracer and find that for strong bath memory the stationary distribution is no longer exponential, and exhibits Gaussian tails. We find the limiting distributions analytically by a time-scale separation approximation on the system.
We also study a non-instantaneous resetting protocol, where the tracer returns to the origin deterministically at a fixed velocity. Unlike the Markovian diffusion, where the stationary state is largely insensitive to the details of the return dynamics~\cite{pal2019invariants,pal2019time}, in this case the return velocity  strongly affects the stationary distribution of the probe. We find an analytic form of this distribution in the slow return regime.

The rest of the paper is organized as follows. In Sec.~\ref{sec:model}, we introduce the viscoelastic setup without stochastic resetting. We discuss the instantaneous resetting protocol in Sec.~\ref{sec_inst}, studying the exact time-dependent moments and the stationary distribution of the tracer in the different parameter regimes. We then study the non-instantaneous resetting protocol in  Sec.~\ref{sec:nonisnt}, by numerically studying the dynamical position fluctuations and stationary distributions, and finding the limiting cases analytical. Finally, in Sec.~\ref{sec:conc}, we conclude by discussing our main results and the possible future directions.

\section{Modeling the viscoelastic bath}
\label{sec:model}
\subsection{Single Bath particle Model}
\label{subsec:bath_part}
\begin{figure}
    \centering
    \includegraphics[width=0.87\linewidth]{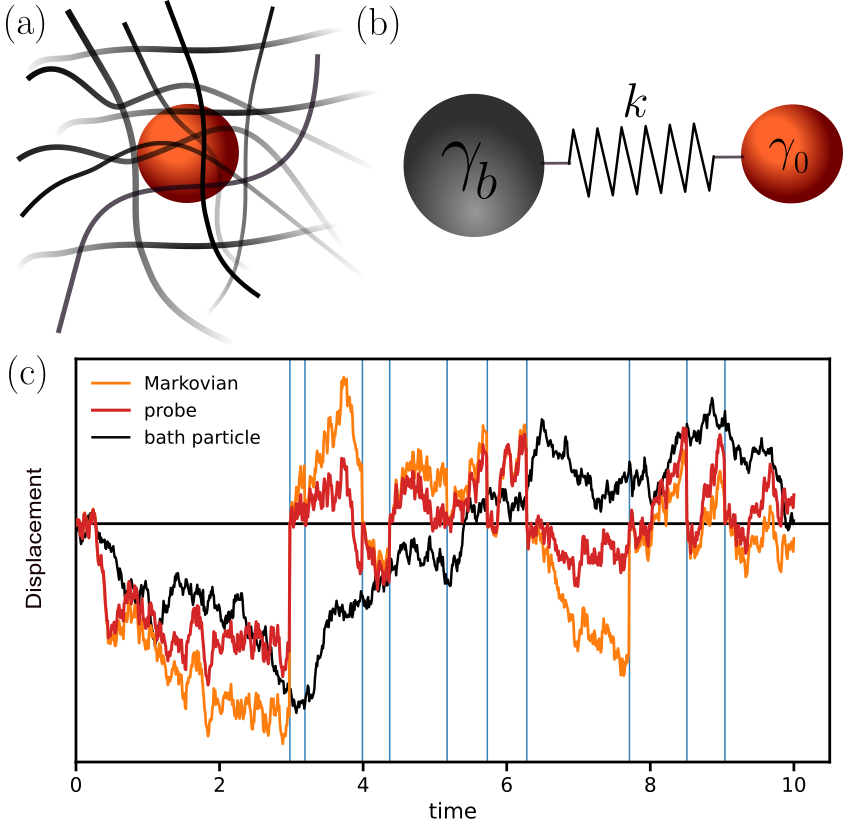}
    \caption{ (a) Schematic of a colloidal probe (in red circle) in a viscoelastic medium, represented by the black mesh. (b) Our  model models the medium by means of another fluctuating particle (in dark grey sphere) coupled to it by a spring (see Eqs.~\eqref{eq:model}). At each resetting event, the position of the probe is reset without affecting the medium directly. (c) Instantaneous Reset: Representative trajectories of the probe for different viscoelastic time-scales in the presence of resetting. }
    \label{fig:model_res}
\end{figure}
Consider the tracer, at position $x(t)$, with bare friction coefficient $\gamma_0$, coupled via a harmonic spring of stiffness $k$ to another particle at position $q(t)$, with bare friction $\gamma_b$( {\it{see}} Fig.~\ref{fig:model_res} (a) and  (b)). The particle $q(t)$ is a toy model for the background viscoelastic medium. The Langevin equations of motion of this coupled system is given by,
\begin{subequations}
\begin{align}
    \gamma_0\dot{x}(t)&=-k(x(t)-q(t))+\sqrt{2} \gamma_0\,\eta_x(t)\\
    \gamma_b\dot{q}(t)&=k(x(t)-q(t))+\sqrt{2\gamma_b}\,\eta_q(t)
\end{align}\label{eq:model}
\end{subequations}
Here, $\eta_x$ and $\eta_q$ are Gaussian random variables with unit variance.

We introduce the dimensionless parameter $\gamma = \gamma_b/\gamma_0$. Further, we write length in the units of $k_BT/k$, time in units of $\gamma_0/k$ and thus have the dimensionless equations of motion,
\begin{subequations}
\begin{align}
    \dot{x}(t)&=-(x(t)-q(t))+\sqrt{2} \eta_x(t)\\
    \gamma\dot{q}(t)&=(x(t)-q(t))+\sqrt{2\gamma} \eta_q(t)
\end{align}\label{eq:model2}
\end{subequations}
  An autonomous time-evolution equation for the tracer particle can be obtained by integrating out the $q$ degree of freedom, leading to,
\begin{align}
\int_{0}^{t}\Gamma(t-t')\dot{x}(t')dt'= F(t).\label{equi:lle}
\end{align}
 The  dissipation kernel $\Gamma(t)$ and noise $F(t)$ are given by,
\begin{align}
\Gamma(t)&=\delta(t)+\gamma^{-1}e^{- t/\gamma},\label{eff:diss:eq}\\
F(t)&=\eta_x(t)+\gamma^{-1}\int_{0}^{t} dt' e^{-(t-t')/\gamma}f(t')\label{eff:noise:eq}.
\end{align}
The dissipation unlike standard Brownian motion depends on a finite-time history of the particle, characterized by the time $\gamma^{-1}$. Such single stochastic bath particles were first introduced as a tool to generate anomalous diffusion in disordered media~\cite{siegle2010markovian,goychuk2012viscoelastic}. Recently, such setups have also been successful in reproducing various experimental observations of tracer particles in viscoelastic fluids~\cite{gomez2015transient,ginot2022barrier,ginot2022recoil}.

Since the process $\mathbf X^T=(x(t)~q(t))$ is Gaussian, The distribution $P(x,q,t)$ is exactly known and has the expression
\begin{align}
    P(x,q,t)=\frac{1}{2\pi\sqrt{\det \Sigma(t)}}
\exp\!\left[
-\frac{1}{2}\mathbf X^{\mathsf T}\Sigma^{-1}(t)\mathbf X
\right],\quad 
\end{align}
 where the elements of the covariance matrix are,
\begin{align}
\Sigma_{11} &=
\frac{\gamma^2\big(1-e^{-2\lambda t}\big)+2(\gamma+1)t}{(\gamma+1)^2}, \label{eq:x2}\\
\Sigma_{12}=\Sigma_{21} &=
\frac{2(\gamma+1)t-\gamma\big(1-e^{-2\lambda t}\big)}{(\gamma+1)^2},\\
\Sigma_{22} &=
\frac{2(\gamma+1)t+\big(1-e^{-2\lambda t}\big)}{(\gamma+1)^2}.
\end{align}
with $\lambda=1+\gamma^{-1}$. This leads to the Gaussian marginal distribution for the position of the tracer particle,
\begin{align}
    P(x,t)=\frac{1}{\sqrt{2\pi\langle x^2(t)\rangle}}
\exp\!\left[-\frac{x^2}{2\langle x^2(t)\rangle}\right].
\end{align}
The particle is diffusive at times much larger than the bath memory time $t\gg \gamma$, with an effective diffusion constant $D_{\mathrm{eff}}=1+\gamma^{-1}$. Thus, the tracer dynamics is Gaussian and diffusive at long times. Do the familiar features of diffusion with stochastic resetting carry over to this non-Markovian viscoelastic setting? We explore this in the following sections.

\begin{figure}
    \centering
    \includegraphics[width=0.84\linewidth]{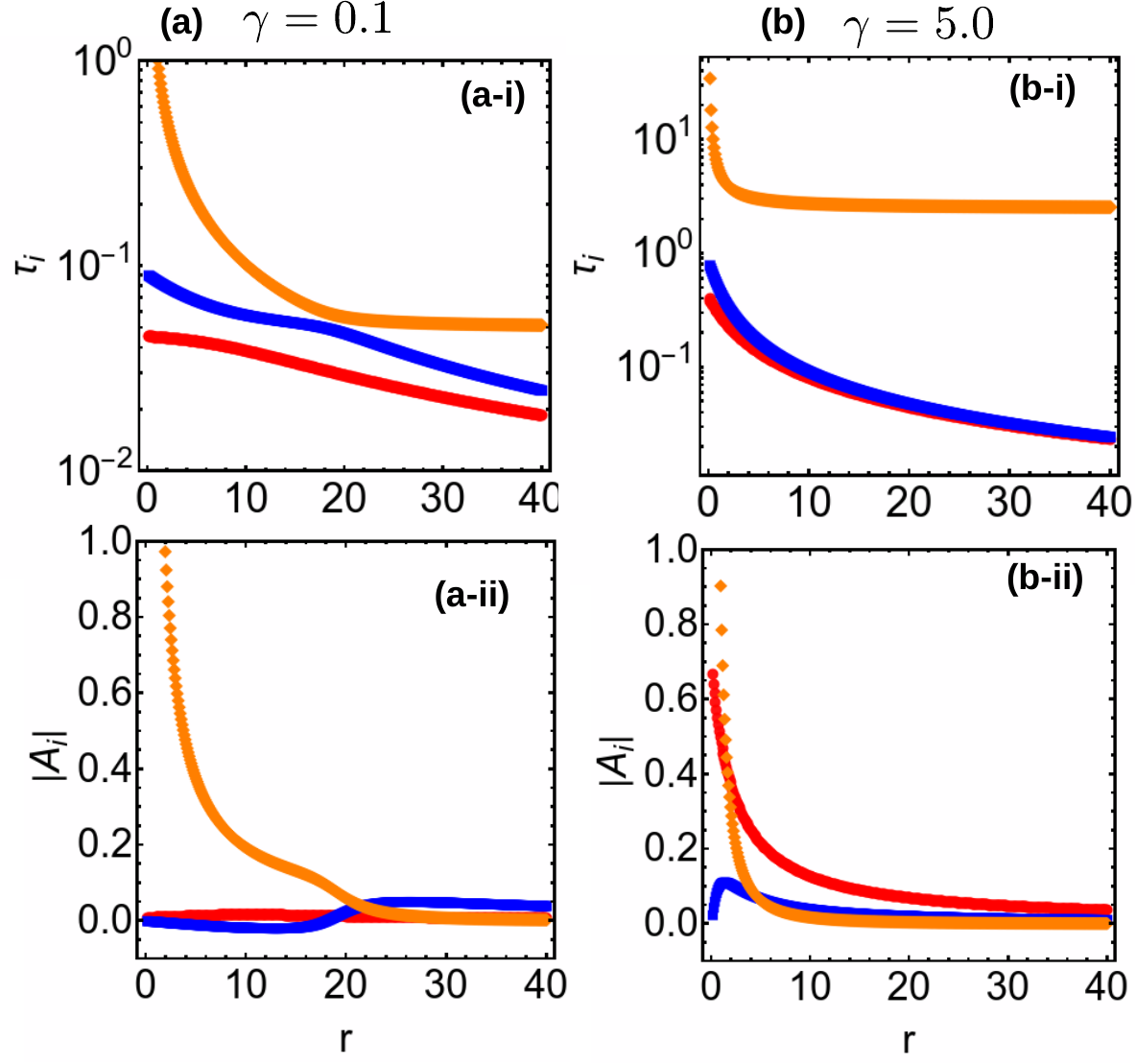}
    \caption{A plot of the extracted time-scales $\tau_i$ and the corresponding weights $A_i$ in Eq.~\eqref{eq:msd} in the small [panels (a)] and large [panels (b)] $\gamma$ limit.}
    \label{fig:timescale}
\end{figure}
\section{Instantaneous reset}
\label{sec_inst}
We now introduce stochastic resetting in our model in the following way.
 At times, chosen from a Poisson distribution $p(\tau)=re^{-r\tau}$, we reset the position $x(t)$ of the tracer  particle to the origin. No direct resetting is applied to $q$, implying that the viscoelastic medium is affected by the resetting only indirectly through the tracer. The resulting dynamics is thus controlled by the two timescales resetting time $r^{-1}$ and the bath relaxation time $\gamma^{-1}$.
In Fig.~\ref{fig:model_res}(c), we present the trajectories of the probe particle and the bath particle in presence of resetting. The trajectory of a particle undergoing stochastic resetting in Markovian bath is also  
presented for comparison.


\label{sec:instpos}
\subsection{Fokker Plank Equations}
\label{sec:FP}
Let $P(x, q, t)$ be the joint probability distribution. The Fokker-Planck equation with resetting of particle 1 at rate \( r \) is:

\begin{align}
    \frac{\partial P}{\partial t} = \mathcal{L}P - r P(x, q, t) + r \delta(x) \int dx' P(x', q, t)\label{eq:fp1}
\end{align}

\noindent
where \( \mathcal{L} \) is the Fokker--Planck operator without resetting:
\begin{align}
\mathcal{L}P = 
\partial x \left[ (x - q) P \right]
+  \partial_x^2 P
+ \gamma^{-1}\partial_q \left[ (q - x) P \right]
+ \gamma^{-1}\partial_q^2 P\nonumber
\end{align}

\subsection{Moments}
To characterize the spatial correlations, we compute the moments of the joint distribution $P(x,q,t)$.  We define the general mixed moments of the system as $M_{l,n}=\langle x^{l} q^{n}\rangle$ (with $l,n>0$).  From the Fokker--Planck equation Eq.~\eqref{eq:fp1}, the evolution equation for the moments reads
\begin{align}
    \dot M_{l,n}&=-(l+\frac n\gamma+r)M_{l,n}+r\delta_{l\,0}M_{0,n}\\
    &+lM_{l-1,n+1}+\frac{n}{\gamma}M_{l+1,n-1}\cr
    &+\frac{n(n-1)}{\gamma}M_{l,n-2}+l(l-1)M_{l-2,n}\nonumber
    \label{eq:momementgen}
\end{align}
with $M_{0,0}=1$.
To obtain the second order moments, we find from Eq.~\eqref{eq:momementgen} a closed linear system of equations,
\begin{subequations} 
\begin{figure*}
        \centering
        \includegraphics[width=0.44\linewidth]{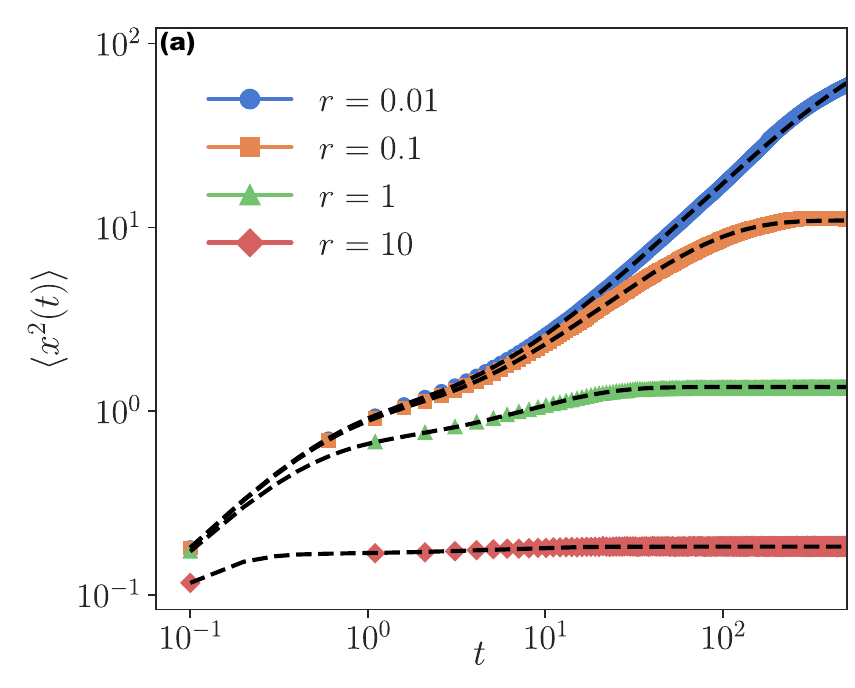}
        \includegraphics[width=0.44\linewidth]{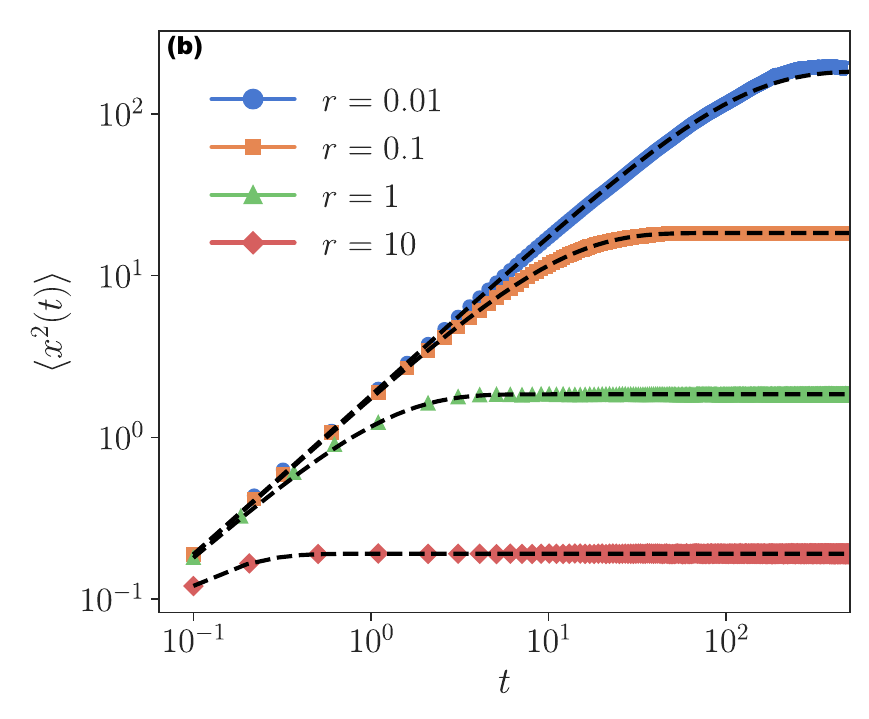}
        \caption{ Mean squared displacement $\la x^2(t)\ra$ of the probe in presence of stochastic resetting. Panel (a) shows the two-step relaxation in the large $\gamma$ limit ($\gamma = 10$ and different values of $r$) as predicted from Eq.~\eqref{eq:msd}. Panel (b) shows $\la x^2(t)\ra$ in the small $\gamma$ limit ($\gamma = 0.1$ and different values of $r$) which has a single step relaxation as predicted. In both the figures the symbols denote numerical simulations while the dashed black lines are obtained from inverting Eq.~\eqref{eq:m2s} for the specific parameter values.}
        \label{fig:msd}
    \end{figure*}

\begin{align}
\dot M_{2,0}  &= -\left(2+r\right) M_{2,0}
+2 M_{1,1}
+ 2, \\
\dot M_{1,1}  &= -\left(\frac{1}{\gamma}+1+r\right) M_{1,1}
+  M_{0,2}
+ \frac{1}{\gamma}M_{2,0}, \\
\dot M_{0,2}  &= -\frac{2}{\gamma} M_{0,2}
+ \frac{2}{\gamma} M_{1,1}
+ \frac{2}{\gamma}.
\end{align}
\label{eqs:2ndmom}
\end{subequations}
We choose the fixed initial condition $x(0)=y(0)=0$ for brevity, and find the second moment of the tracer position in Laplace space,
\begin{align}
    \tilde{M}_{2,0}(s)&=\frac {2 (\gamma s + 2) (\gamma (s' + 1) +  1)} {s\left (\gamma^2 s (s' + 1) (s'+ 2) + s'[\gamma  (2 s' +  s + 4) + 2 ] \right)}\label{eq:m2s}
\end{align}
with the notation $s'=s+r$ for brevity. The pole at $s=0$ indicates a long-time stationary value, expected due to the resetting. Let us first analyse how the relaxation to this stationary value occurs. To this end, we rewrite Eq.~\eqref{eq:m2s} as
\begin{align}
     \tilde{M}_{2,0}(s)=\frac{M^s_{2,0}}{s}-\sum_{i=1}^3\frac{A_i}{s-s_i}
\end{align}
where $M^s_{2,0}$ is the stationary value of the second moment, $s_i$ are the non-zero poles of the denominator, and $\{A_i\}$s are the residues of $\tilde{M}_{2,0}(s)$ at the $i$th pole. This implies the form,
\begin{align}
    M_{2,0}(t)=M^s_{2,0}+\sum A_i e^{-t/\tau_i}\label{eq:msd}
\end{align}
in the time domain, where the relaxation time-scales are given by the inverse of the real part of $\{s_i\}$. Clearly, the multi-step decay is visible when the weights $A_i$ of the faster decay modes are large. Exact closed form solutions of $\tau_i$ and $A_i$ are complicated and and provide little practical insight. Here, we can obtain them numerically, and understand at which parameter regimes the multi-step decay is visible. Figure.~\ref{fig:timescale} shows the $\{A_i,\tau_i\}$ pair for small and large values of $\gamma$, as a function of the resetting rate.  We see that for small $\gamma$, the $A_i$s corresponding to the faster time-scales are always small, hence $M_{2,0}(t)$ looks like a single exponential growth to the saturation value. 
On the other hand, for large $\gamma$, one of the faster time-scales has a large $A_i$, which leads to a two step exponential decay with a plateau at $t\sim\tau_1$.

These are confirmed from looking at $M_{2,0}(t)$ in these regimes. In Fig.~\ref{fig:msd} we invert Eq.~\eqref{eq:m2s} for specific values of $(\gamma,r)$ showing the above discussed qualitative features, and compare them with numerical simulations, which show very nice agreement. 

\begin{figure*}
    \centering
    \includegraphics[width=0.44\linewidth]{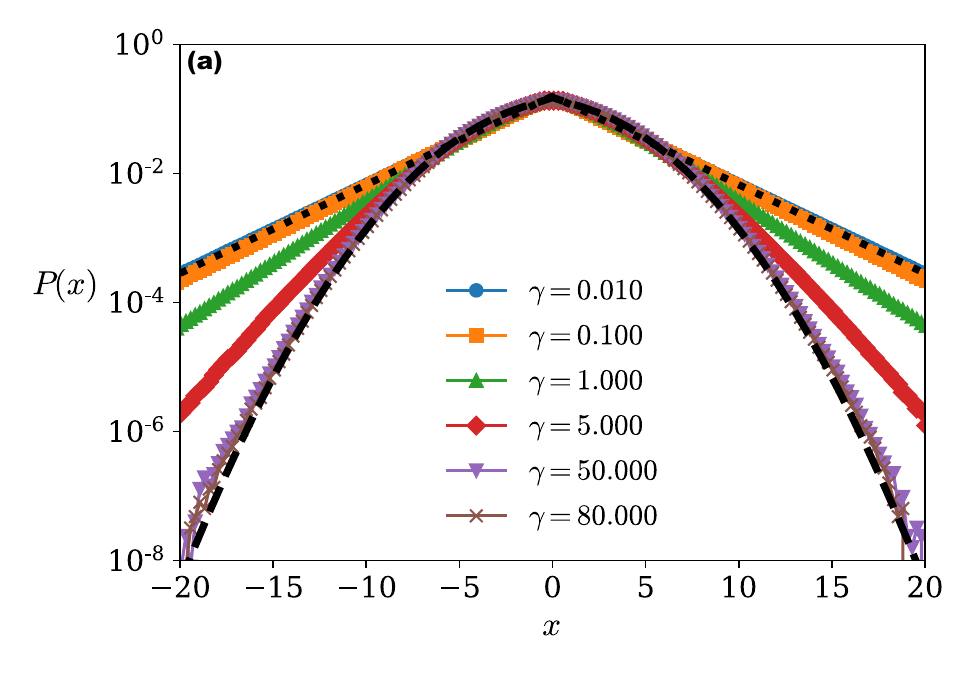}
     \includegraphics[width=0.44\linewidth]{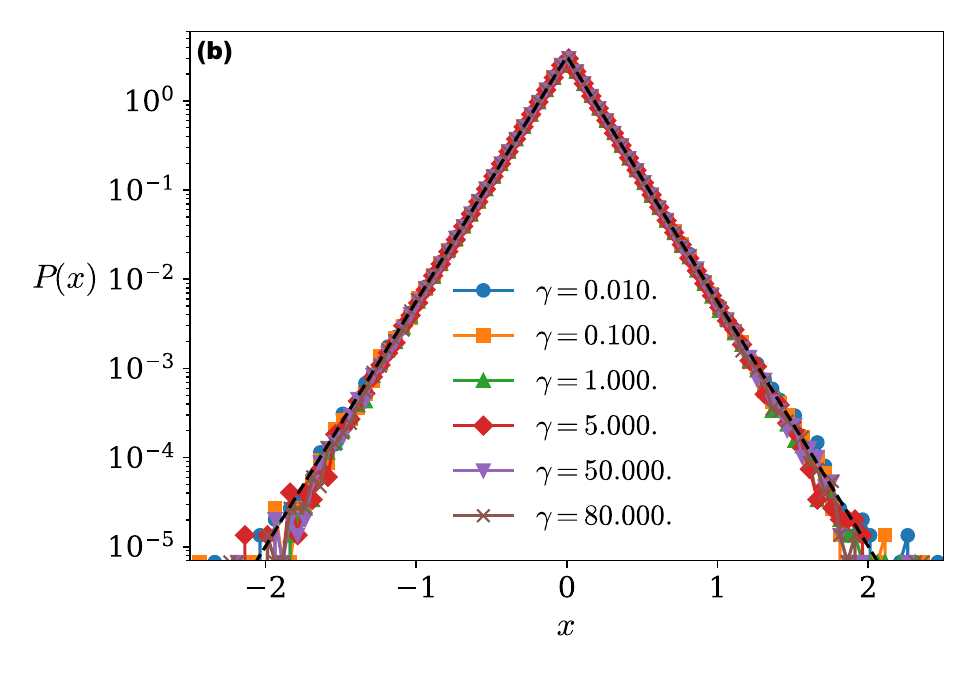}
    \caption{Stationary position distribution $P(x)$ of the probe under instantaneous resetting for different viscoelastic time-scales $\gamma$.Panel (a) shows $P(x)$ for small resetting rates and different values of $\gamma$.The dotted black lines denote the Markovian Brownian motion limit Eq.~\eqref{eq:BM}, while the dashed line denotes the analytical prediction Eq.~\eqref{int:psx} for large $\gamma$. Panel (b) shows the independence of $P(x)$ with $\gamma$ for large resetting rates, as predicted in Eq.~\eqref{eq:markovreset}.}
    \label{fig:instdisttt}
\end{figure*}

This can be physically understood in the following way. The dynamics of the tracer is governed by the competition between three characteristic time-scales: the mean reset time $\tau_r = r^{-1}$, the viscoelastic relaxation time of the bath $\tau_b = \gamma$, and the equilibration time of the relative tracer-bath coordinate $z = x - q$. The latter follows an Ornstein-Uhlenbeck process,
\begin{align}
    \dot z=-\frac{z}{\tau_z}+\zeta\quad\text{with }\tau_z=\frac{\gamma}{1+\gamma}
\end{align}
where $\zeta$ is a zero mean Gaussian white noise. The appearance of multiple decay modes reflects the separation between these underlying dynamical processes. It is important to note that the relaxation times appearing in the second moment dynamics  (see Eq.~\eqref{eq:msd}) do not coincide directly with these microscopic time-scales. Instead, they correspond to the eigenmode relaxation times of the coupled moment equations arising from the interplay between resetting, tracer-bath equilibration, and bath relaxation.

The limit $\gamma\to 0$ is actually the Markovian limit, where the bath relaxes rapidly and effectively follows the tracer adiabatically, resulting in a single visible relaxation time. On the other hand, for large gamma, first the tracer rapidly equilibrates with respect to the quasi-static bath configuration, producing a fast initial growth of $M_2(t)$ and a transient plateau.
Subsequently, the slow relaxation of the bath, on the timescale $\sim\tau_b$, drags the tracer towards the final stationary state, leading to a second exponential growth. 

Now let us analyse the stationary value of the position fluctuations,
\begin{align}
    M^s_{2,0}=\frac {4 (\gamma (r + 1) + 1)} {\gamma (2 r + 4) r + 2 r}.\label{eq:instmsd}
\end{align}
In the limit $\gamma\to 0$ goes to $M^s_{2,0}$ reduces to the Markovian Brownian motion result $2/r$, while for $\gamma\to \infty$ it approaches $2(r+1)/(r^2+2r)$.

 The structure of the equations in Eq.~\eqref{eq:momementgen} allow us to write any of the general higher order correlation in terms of the lower order correlations. The next non-zero moment $\la x^4\ra$ can also be computed, using Eq.~\eqref{eq:momementgen}. Using this we compute the excess kurtosis, $\kappa=(\la x^4\ra-3\la x^2\ra)/\la x^2\ra$, which is a measure of the tailedness of the position distribution. The exact expression is rather long and presented in the Appendix, we present some interesting limits: (i) $\lim_{\gamma\to 0}\kappa=3$ for any $r$; (ii) $\lim_{r\to \infty}\kappa=3$ for any $\gamma$; (iii)  $\lim_{\gamma\to\infty}\kappa$ for small value of $r$ is close to zero; and the trivial no resetting limit $\lim_{r\to 0}\kappa=0$. These are demonstrated in Fig.~\ref{fig:kurtosis}, and we will discuss them when discussing the shape of the stationary distribution.
 
\subsection{Stationary Distribution}

In this section, we analyze the stationary distribution $P(x)$ that emerges due to resetting in the viscoelastic fluid. In Fig.~\ref{fig:instdisttt}(a), we have plotted the numerically obtained position distribution for different values of $\gamma$.   Obtaining an exact analytical expression for $P(x)$ from Eq.~\eqref{eq:fp1} for general values of the parameters $\gamma$ and $r$ is not straightforward. However, the exact form of the distribution can be obtained in several interesting limiting regimes.

For any $r$, $\gamma\to 0$ is the limit of Markovian Brownian motion, 
\begin{align}
P(x) =   \frac{\sqrt{r}}{2}e^{-\sqrt{r}|x|}.
\label{eq:BM}
\end{align}
This is consistent with the limit (i) of the kurtosis discussion earlier. It is interesting to note that the excess kurtosis is always bounded withing $[0,3]$, indicating that the tails are not slower than an exponential, and not faster than a Gaussian. 

In the large $\gamma$ limit, the bath is almost static. Thus, we can treat the tracer dynamics in a quasistatic limit of the bath, i.e., making a time-scale separation in the Fokker-Planck equation Eq.~\eqref{eq:fp1} with $q$ as a constant parameter. The distribution of the tracer given a fixed position $q$ of the bath is the solution of,
\begin{align}
   \partial_t P_q= \partial x \left[ (x - q) P_q \right]
+  \partial_x^2 P_{\text{q}}-r P_q + r \delta(x) \int dx' P_q(x', q, t).\label{eq:probe-quasistaticbath}
\end{align}
This suggests that the tracer dynamics for a quasistatic bath is an Ornstein-Uhlenbeck particle with the harmonic trap centered at $q$, in the presence of random positional resets to the origin. The slower bath dynamics in this limit can be obtained as,
\begin{align}
    \gamma\dot{q}(t)=-\la q-x\ra_q+\sqrt{2\gamma}\eta_q(t)\label{eq:bathqs}
\end{align}
To compute the first term on the rhs, it is sufficient to compute $\la x\ra_q$. To this end, we note that the mean of an OU particle (with unit diffusion coefficient, and unit trap strength) centered at $q$ is given by $q(1-e^{-t})$. In the presence of resetting the desired quasistatic mean is given by,
\begin{align}
    \la x\ra_{\text{q}}=r\int_0^\infty dt'e^{-rt'}q(1-e^{-t'})=\frac{q}{1+r}.
\end{align}
This leads to the effective bath dynamics from Eq.~\eqref{eq:bathqs},
\begin{align}
    \gamma\dot{q}(t)=-\frac{r}{1+r}q(t)+\sqrt{2\gamma}\eta_q(t).
\end{align}
Thus the stationary bath distribution $\rho(q)$ is a Gaussian,
\begin{align}
    \rho(q)=\frac{e^{-\frac{rx^2}{1+r}}}{\sqrt{2\pi(1+r)/r}}.
    \label{eq:bath}
\end{align}
 Thus we can treat the probe dynamics as a resetting OU particle with its center $q$ distributed according to $\rho(q)$. The stationary distribution of the resetting tracer in this limit is then, 
\begin{align}
    P(x)&=\int_{-\infty}^{\infty} dq\,\rho(q)\left[r\int_0^{\infty} dt' e^{-rt'}  P_q(x,t)\right]
    \end{align}
    where $P_q(x,t)$ is the solution of Eq.~\eqref{eq:probe-quasistaticbath} for $r=0$, i.e., the Ornstein--Uhlenbeck propagator with trap center fixed at $q$,
\begin{align}
P_q(x,t)=\frac{1}{\sqrt{2\pi(1-e^{-2t})}}
\exp\left[-\frac{\left(x-q(1-e^{-t})\right)^2}{2(1-e^{-2t})}\right].
\end{align}
Averaging this propagator over the quasistatic bath distribution $\rho(q)$ in Eq.~\eqref{eq:bath}, we obtain
    \begin{align}
 P(x) &=r\int_0^{\infty} dt' e^{-rt'}\frac{1}{\sqrt{2\pi \sigma^2(t)}}e^{-\frac{x^2}{2\sigma^2(t)}}\label{int:psx}
\end{align}
where  $\sigma^2(t)=(1-e^{-2t})+\frac{1+r}{r}(1-e^{-t})^2$. It is difficult to obtain an exact closed form solution of the above integral, however, it can be numerically evaluated. This is compared to numerical simulations in Fig.~\ref{fig:instdisttt} (a) and shows excellent agreement. 

\begin{figure}
    \centering
    \includegraphics[width=0.89\linewidth]{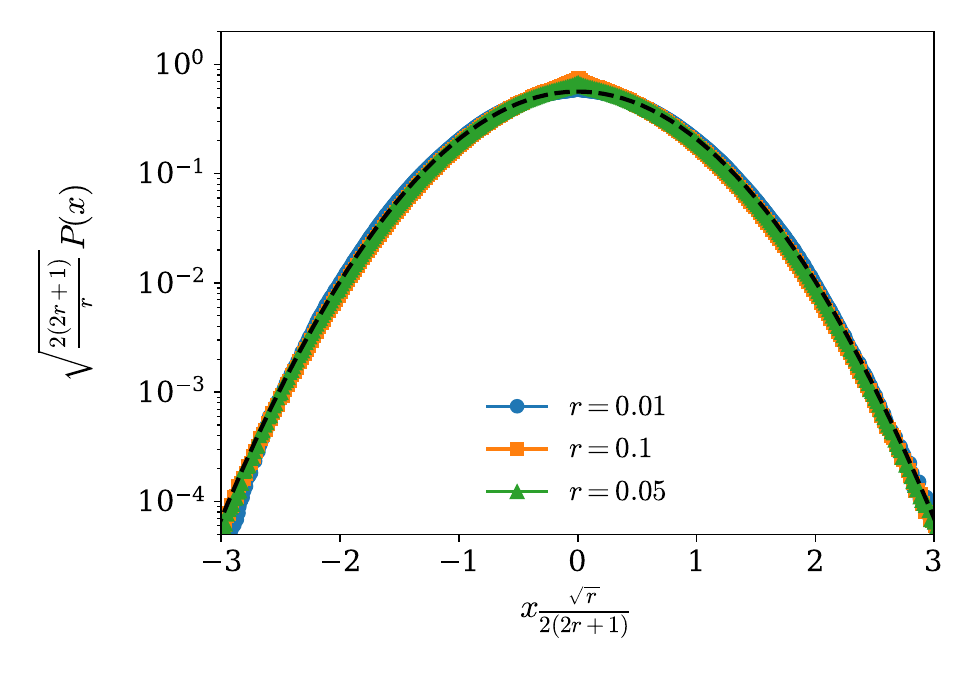}
\caption{Assymptotic behavior of the stationary position distribution at the tails and near the origin respectively. Panel (a) shows the scaled position distribution, the scaling prediction in Eq.~\eqref{eq:scaling} is illustrated by a collapse of the distributions obtained numerically for different values of $r$, and $\gamma=100$. The symbols denote the numerical simulations, while Gaussian scaling function is shown with dashed black line.   Panel (b) shows the behavior of the distribution near the origin which shows an exponential decay  as predicted in Eq.~\eqref{eq:scaling} (shown as a black dashed line).}
    \label{largegam:assymptotes}
\end{figure}

We can, however, get some asymptotic forms of the stationary distribution from Eq.~\eqref{int:psx}. The integrand in Eq.~\eqref{int:psx} is Gaussian with a bounded width, thus, for large $x$, the contribution to the integral is dominated by the largest value of $\sigma^2(t)$, i.e., $\sigma^2(t\to\infty)=(2+r)/r$. This gives the tails of the stationary distribution,
\begin{align}
    P(|x|\to\infty)=\frac{1}{\sqrt{2\pi(2r+1)/r}}e^{-\frac{rx^2}{2(2r+1)}}.\label{eq:scaling}
\end{align}
Thus we see that the tails become Gaussian. 
This asymptotic behavior of the tails is compared with numerical simulations in Fig.~\ref{largegam:assymptotes} and show good agreement.

In the limit of fast resetting $r\gg 1$, on the other hand, the tracer experiences resetting at a much faster time-scale than it can experience the interactions with the bath. Due to this the tracer behaves like a Markovian diffusion in the presence of resetting. This limit is illustrated in Fig.~\ref{fig:instdisttt} (b), which shows that stationary distribution in this limit is independent of $\gamma$, and given by 
\begin{align}
P(x)=\frac{\sqrt{r}}{2}e^{-\sqrt{r}|x|}. \label{eq:markovreset}
\end{align}

\section{Non-instantaneous resetting}
\label{sec:nonisnt}
\subsection{Return protocol}

In the earlier sections, we assumed that resetting occurs instantaneously, i.e., the diffusing particle is effectively teleported back to the origin and immediately resumes its motion. In realistic situations, however, resetting  another requires a finite amount of time, and this return dynamics must be explicitly taken into account to characterize the non-equilibrium states. Experimental realizations of stochastic resetting are therefore inherently non-instantaneous, and several return protocols have been proposed~\cite{pal2019time}. 

 Here, we consider stochastic resetting with a finite return speed. The tracer diffuses according to the dynamics specified in Eq.~\eqref{eq:model2}. At a constant reset rate  $r$, it undergoes a reset event and switches to the return phase. In the return phase, the tracer moves deterministically toward the origin with a constant speed. Once it reaches the resetting position, it resumes its diffusive dynamics. Throughout the notion, the bath particle $q(t)$ continues to evolve diffusively and remains linearly coupled to $x(t)$.
The retrun phase dynamics thus follows
\begin{subequations}
\begin{align}
    \dot{x}(t) &= -\mathrm{sgn}[x(t)]\, v_0, \\
    \gamma \dot{q}(t) &= (x(t)-q(t)) + \sqrt{2\gamma}\,\eta_q(t),
\end{align}
\end{subequations}
\begin{figure}
    \centering
    \includegraphics[width=0.88\linewidth]{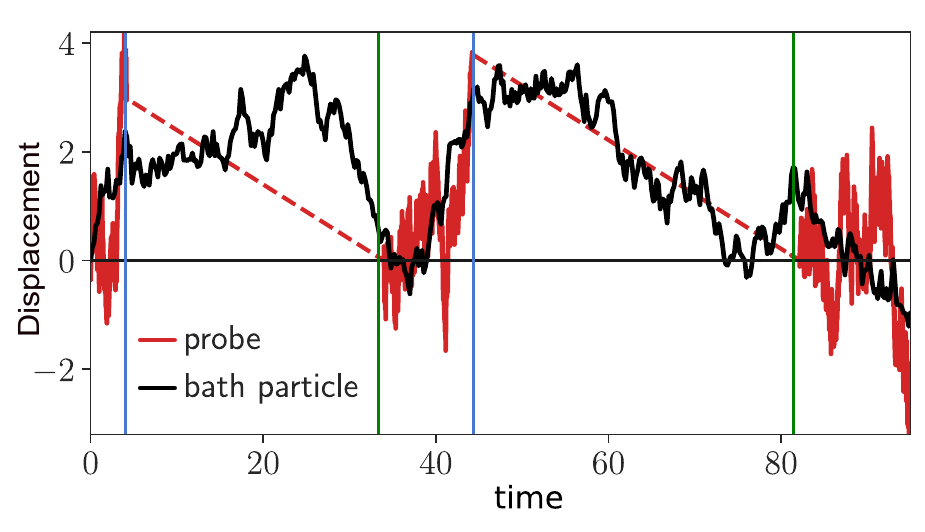}
    \caption{Non Instantaneous Reset: Representative trajectories of the probe (red) and the bath particle(black) evolving via non-instantaneous reset. The onset of the return phase is indicated by the blue line when the probe undergoes a ballistic dynamics to reach $x=0$ marked by the green vertical line. Note that the bath particle evolves diffusively in the entire region.}
    \label{fig:traj_non}
\end{figure}
The dynamics, along with the return protocol, are illustrated in Fig.~\ref{fig:traj_non}. 

\subsection{Moments}
\label{sec:noninst-mom}
\begin{figure}
    \centering
        \includegraphics[width=0.78\linewidth]{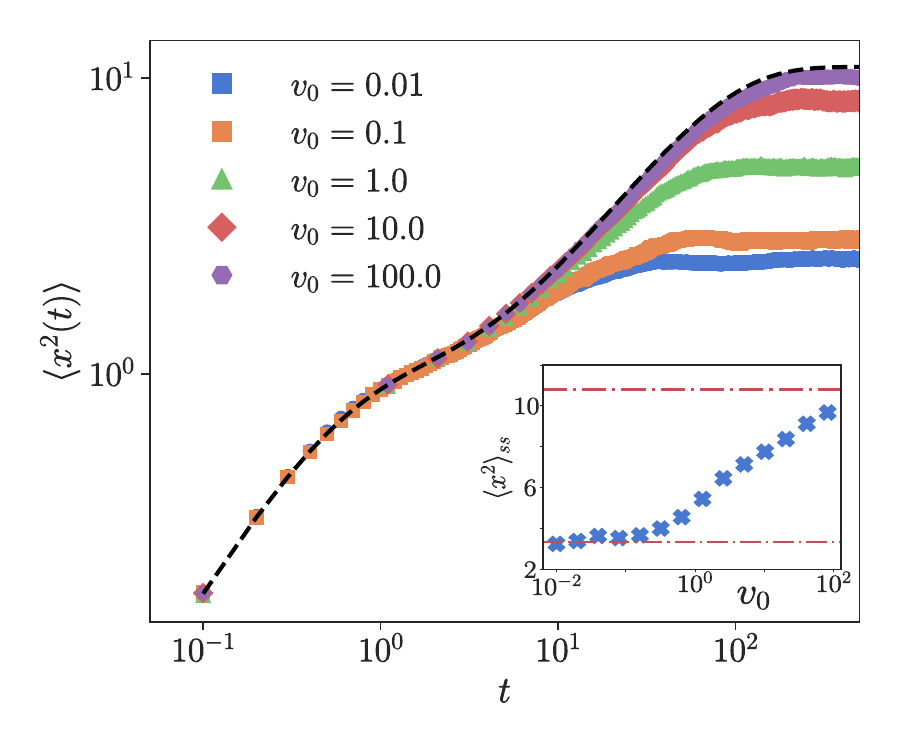}
\caption{Mean square displacement for different return speed $v_0$ keeping $\gamma = 10$. The msds display a plethora of non equilibrium steady states defined by $v$ dependent long time saturation values $\langle x^2\rangle_{ss}$ for $r = 0.1$.  The black dotted lines denote the msd from instantaneous reset obtained theoretically. The {\it {Inset}} shows the long time saturation value of the MSD $\langle x^2 \rangle_{ss}$ as a function of the return speed $v_0$. The red dash-dot  lines  denote the large and small $v_0$ asymptotes Eq.~\eqref{eq:instmsd} and \eqref{eq:noninst:msd}, respectively}
    \label{fig:msd_v}
\end{figure}

Here, we discuss how the introduction of a finite return speed $v_0$ modifies the  position fluctuations $\langle x^2(t)\rangle$ of the tracer. Unlike instantaneous resetting, the resulting dynamics is not amenable to a full analytical treatment. We therefore present numerical simulation results that capture the dominant physical features and analyze several limiting regimes to gain further insight. The mean squared displacement of the tracer is displayed in Fig~\ref{fig:msd_v}.

We first consider the two known Markovian limit: (i) For small values of $\gamma \to 0$, the bath relaxes rapidly compared to the tracer and the relaxation toward the stationary state proceeds in a single step, similar to the instantaneous resetting case~\cite{pal2017first}
The other Markovian limit  is give by (ii) large resetting rates ($r \gg 1$). Here, the tracer experiences reset events on timescales much shorter than the viscoelastic relaxation time. In this regime the system does not have sufficient time to build up correlations with the bath between successive resets, and the dynamics effectively becomes Markovian. As a result, both the time-dependent fluctuations and the stationary distribution become independent of $\gamma$ and insensitive to the details of the return protocol, recovering the exponential stationary form characteristic of Brownian motion with resetting. 

For large $\gamma$, where viscoelastic memory is strong, the characteristic two-step relaxation persists characterized an initial rapid growth followed by a second relaxation towards stationarity. Thus, the qualitative structure of the relaxation dynamics remains unchanged upon introducing a finite return speed. In fact, the relaxation time also does not change due to the return velocity. This is because the return phase for the tracer is deterministic, and the relaxation dynamics of the fluctuations are governed by the diffusive phase of the coupled tracer-bath system, which remains unchanged.

The effect of $v_0$  is reflected in its stationary value ({\it see} Fig.~\ref{fig:msd_v}). For fixed $(\gamma,r)$, the limit $v_0 \to \infty$ is trivial the return time vanishes and we recover the instantaneous resetting result obtained in Sec.~\ref{sec_inst}, as expected. This is also demonstrated in Fig.~\ref{fig:msd_v}. As $v_0$ is decreased further, the stationary variance $\langle x^2\rangle_{ss}$ decreases monotonically approaches a limiting value, as observed in Fig.~\ref{fig:msd_v}(b). This is because small $v_0$ leads to long return times; and allowing the bath also to be close to the origin at the end of each return phase. We discuss this more rigorously along with the corresponding distributions in Sec.~\ref{sec:noninst-dist}.   The limiting stationary variance as $v_0\to 0$ can also be computed and is given in detail in Eq.~\eqref{eq:noninst:msd}({\it see} Appendix B). In Fig~\ref{fig:msd_v}({\it Inset}), we have plotted the long time stationary value as a function of the return speed $v_0$ along with the analytically obtained limiting values for $v_0 \to 0$ and $v_0 \to \infty$.

The dependence of the variance on $v_0$ indicates that the invarinace in the position fluctuations of a resetting Markovian Brownian particle on the return speed is not valid here anymore.

\subsection{Stationary distribution}
\label{sec:noninst-dist}
\begin{figure}
    \centering
    \includegraphics[width=0.79\linewidth]{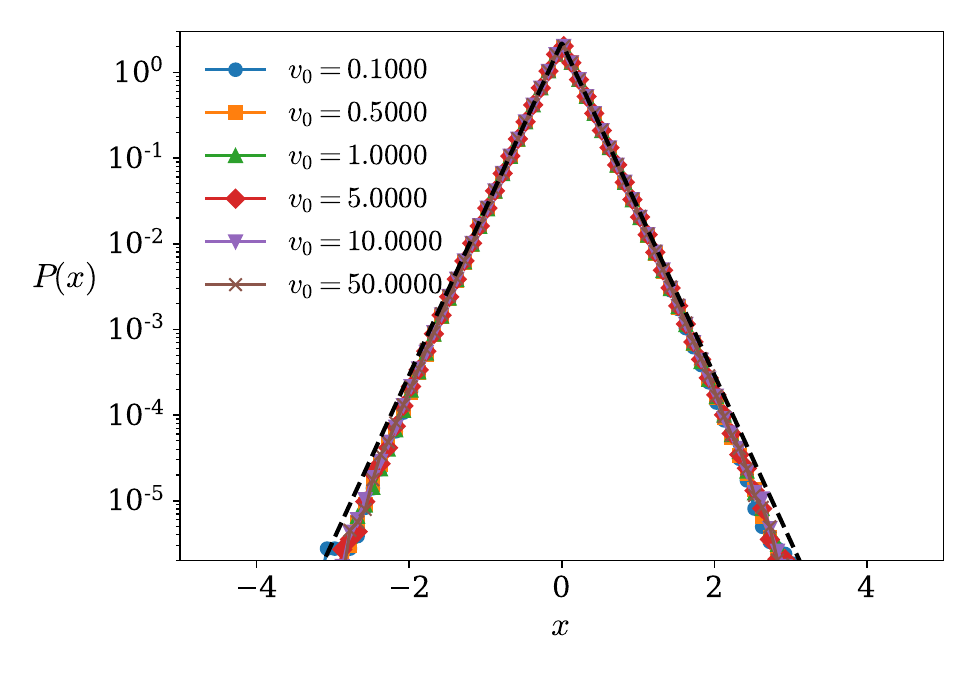}
    \caption{Stationary position distribution of tracer for the non-instantaneous resetting protocol for different values of the return velocity $v_0$, $r=20$, and $\gamma=40$. The colored line-points denote data obtained from numerical simulations, the dashed black line corresponds to Eq.~\eqref{eq:markovreset}. }
    \label{fig:noninstlarger}
\end{figure}

We now discuss how the constant--velocity return protocol affects the stationary position distribution $P(x)$ of the tracer. In overdamped Markovian diffusion, it is known that the full time-dependent density is invariant under constant-velocity return protocols, and it coincides with the result for instantaneous resetting \cite{pal2019time}. As already indicated by the second moment, in the non-Markovian setting this invariance does not hold in general except for specific limits, namely, (i) the bare Markovian limit $\gamma\to 0$, (ii) Large resetting $r\gg 1$ where there is an effective Markovian behavior. The effective Markovian limit in the latter case is due to resetting being so frequent hat the tracer does not have time to build correlations with the bath between successive reset events, and we get Eq.~\eqref{eq:markovreset} as the full stationary tracer distribution. This is illustrated in Fig.~\ref{fig:noninstlarger}, where the stationary distribution converges to Eq.~\eqref{eq:markovreset}.

\begin{figure}
    \centering
    \includegraphics[width=0.79\linewidth]{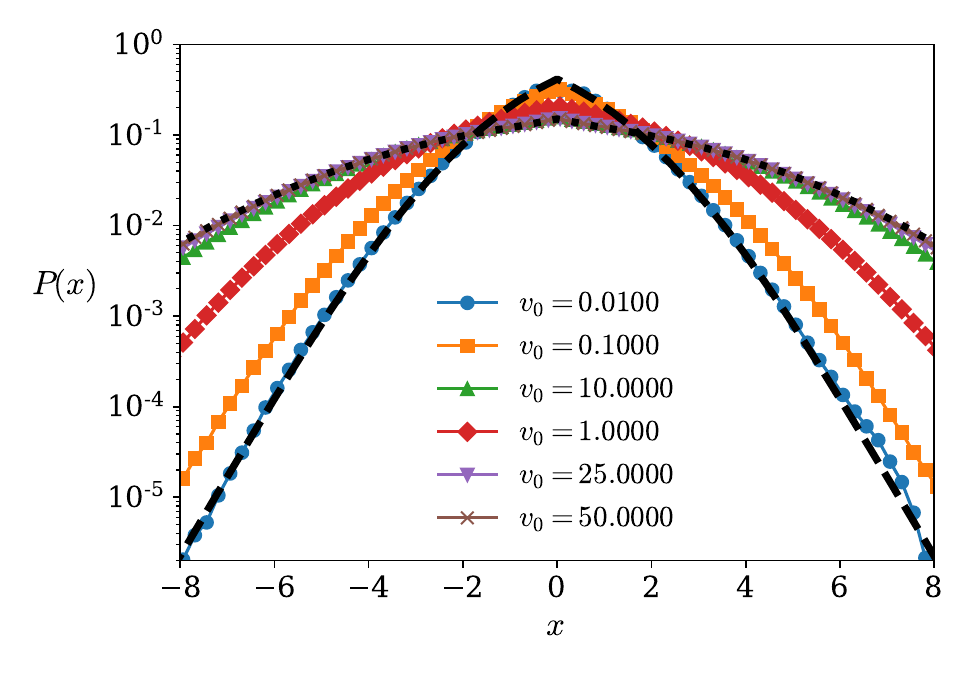}
    \caption{Stationary position distribution of tracer for the non-instantaneous resetting protocol for different values of the return velocity $v_0$, $r=0.1$, and $\gamma=40$. The colored line-points denote data obtained from numerical simulations, the dashed black line corresponds to the predicted distribution Eq.~\eqref{eq:fulldist0} for small $v_0$.}
    \label{fig:fulldistnoninst}
\end{figure}

\begin{figure}
    \centering
    \includegraphics[width=0.79\linewidth]{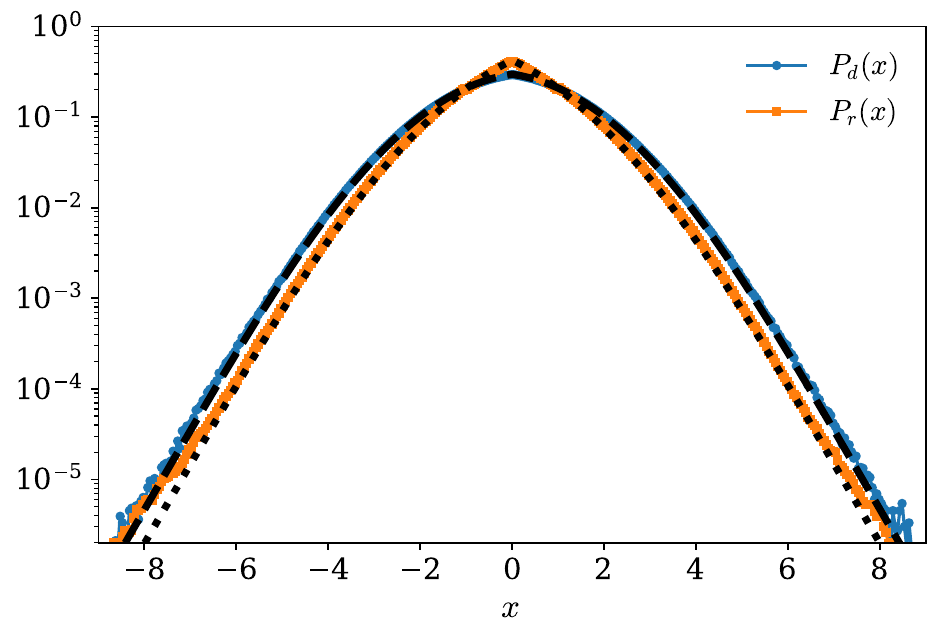}
    \caption{Position distribution of the tracer in the diffusive and return phases in the limit of large $\gamma$, and small $v_0$. The symbols denote numerical simulations, while the dashed, and dotted lines denote the analytical predictions Eqs.~\eqref{eq:diffdist} and \eqref{eq:returndist} for the diffusive and return phases, respectively. For both the curves $\gamma=40$, $r=0.1$, and $v_0=0.01$.}
    \label{fig:diff-return}
\end{figure}

The most interesting regime is where the memory is strong, namely large $\gamma$ at finite resetting rate $r=\mathcal{O}(1)$. In this regime, the stationary distribution depends sensitively on the return speed $v_0$, and protocol invariance breaks down. This is directly visible in Fig.~\ref{fig:diff-return}, while the overall shape of the distribution remains qualitatively similar to the instantaneous case, the distribution becomes progressively narrower as $v_0$ is decreased, with a corresponding reduction of the stationary variance, consistent with the behavior of $\langle x^2\rangle_{ss}$ discussed in Sec.~\ref{sec:noninst-mom}. In the opposite limit $v_0\to\infty$, the return time vanishes and we recover the instantaneous-reset distribution Eq.~\eqref{int:psx}, as demonstrated in Fig.~\ref{fig:fulldistnoninst}.

\begin{figure}
    \centering
    \includegraphics[width=0.79\linewidth]{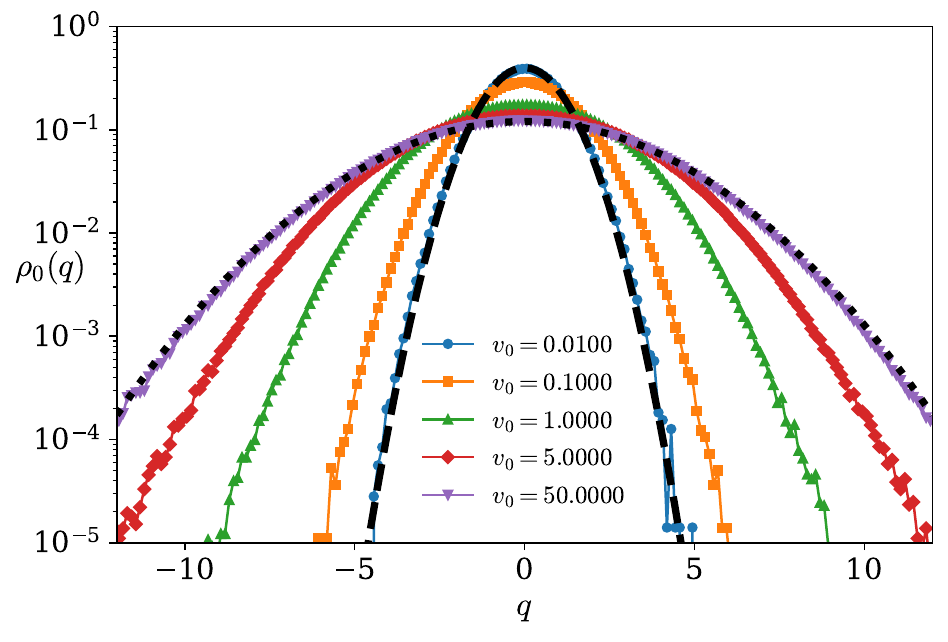}
    \caption{Bath distribution at the end of a resetting event for different values of the return velocity and $\gamma=40$, $r=0.1$. The black dotted and dashed lines denote the analytical prediction for asymptotically large and small return protocols [see Eqs.~\eqref{eq:bath}and \eqref{eq:rhoq2}, respectively].}
    \label{fig:bathatreset}
\end{figure}

The physical origin of this protocol dependence can be understood in the following way. In Markovian diffusion, the statistics of a diffusion excursion following a reset depends only on the reset position [origin in our case]. Equivalently, the propagator has the form $P(x,t|x(0)=0)$, and the return protocol does not influence the statistics of subsequent excursions once the particle is at the origin. In our case, the state of the system at the beginning of a diffusion excursion is specified by the propagator $P(x,t|0,q_0)$, which depends on the bath configuration $q_0$ at the completion of the resetting event. So the position distribution depends crucially on the distribution of $q_0$, which we denote as $\rho_0(q_0)$. 
When $v_0\to\infty$, $\rho(q_0)$ does not evolve during the return phase, and we get the same physical description as in Sec.~\ref{sec_inst}. However, as $v_0$ keeps decreasing, the bath gets the opportunity to equilibriate with the tracer, changing the distribution of the former. This is shown in Fig.~\ref{fig:bathatreset} where $\rho_0(q_0)$ is plotted for different return speeds. 

For $v_0\to \infty$, we get Eq.~\eqref{eq:bath}, as expected [verified with numerical simulations in Eq.~\eqref{fig:bathatreset}]. As the return gets slower, the bath gets more time to relax toward the origin before the tracer reaches the origin and restarts the diffusion phase. This reduces the typical distance $z=x-q$ between tracer and bath at the start of each diffusion segment, and since the tracer dynamics is driven by the $z$, it suppresses the subsequent spreading, leading to a narrower distribution. This becomes particularly useful in an extreme slow-return limit $\frac{|x_0|}{v_0}\gg \gamma$. In this regime, during the return phase the tracer stays close enough to the origin for long enough that the bath relaxes essentially to its own local stationary state around $x\simeq 0$ before the diffusive phase begins. Correspondingly, the bath dynamics during the return phase becomes well-approximated by the decoupled Ornstein--Uhlenbeck relaxation,
\begin{equation}
\dot q(t)=-\gamma^{-1} q(t)+\sqrt{\frac{2}{\gamma}}\,\eta(t),
\end{equation}
which yields a Gaussian stationary distribution
\begin{equation}
\rho_0(q_0)=\frac{1}{\sqrt{2\pi}}\,e^{-q_0^2/2},\label{eq:rhoq2}
\end{equation}
This prediction is confirmed by numerical simulations and illustrated in Fig.~\ref{fig:bathatreset}. 

Thus to obtain the stationary distribution, we can forget about the return phase, and consider that the diffusive phase always has the intial condition $\delta(x)\rho_0(q)$, where $\rho_0$ is a Gaussian. Since the dynamics is linear and Gaussian, the propagator of the tracer position starting from $x(0)=0$ and $q(0)=q_0$ is Gaussian with zero mean (after averaging over $\rho_0(q_0)$) and variance
\begin{equation}
\sigma_d^2(t)=\frac{\gamma^2\big(1-e^{-2\lambda t}\big)+2(\gamma+1)t}{(\gamma+1)^2}+\frac{(1-e^{-\lambda t})^2}{\lambda^2},
\end{equation}
where $\lambda=(1+\gamma)/\gamma$. Thus the stationary distribution in the diffusive phase is,
\begin{align}
    P_d(x)= \int_0^\infty dt\,
r e^{-rt}
\frac{1}{\sqrt{2\pi \sigma_d^2(t)}}
\exp\!\left(-\frac{x^2}{2\sigma_d^2(t)}\right)\label{eq:diffdist}
\end{align}
This integral is again difficult to perform numerically, however, can be evaluated numerically, and agrees with the numerical simulations very well.

During the return phase, the tracer moves deterministically toward the origin following
\begin{equation}
x(t)=x_0-\text{sgn}(x_0)v_0 t
\end{equation}
and reaches the origin after a time
\begin{equation}
T=\frac{|x_0|}{v_0}.
\end{equation}
where the initial point of the return phase $x_0$ is sampled from the stationary distribution in the diffusive phase Eq.~\eqref{eq:diffdist}. The distribution in the return phase is thus,
\begin{align}
    P_r(x)=\frac{1}{\la T\ra_{x_0}}\left\la\int_0^T\delta(x-x_0+\text{sgn}(x_0)v_0 t) dt\right\ra_{x_0}
\end{align}
Since in the return phase, $|x|\leq x_0$, 
\begin{align}
    P_r(x)&=\frac{1}{\la |x_0|\ra_{x_0}}\int_{|x|}^\infty P_d(x')dx' \\
    &=\frac{1}{2\la |x_0|\ra_{x_0}}\int_0^{\infty}dt\,re^{-rt}\text{Erfc}\left(\frac{x}{\sqrt{2\sigma_d^2(t)}}\right).\label{eq:returndist}
\end{align} 
where we used Eq.~\ref{eq:diffdist}, and changed the order of the integration. Again exact closed form expressions for the above integral is difficult, but we can perform the integral numerically. This is compared to simulations in Fig.~\ref{fig:diff-return}, and show good agreement. It is important to point out that the distribution of the return times thus follow a distribution,
\begin{align}
    \mathcal{T}(\tau)=2v_0\,P_d(v_0\tau),\quad \tau>0
\end{align}
which is non-exponential.

The total stationary distribution of the tracer position is then just a sum of the weighted average of the diffusive and return phase distributions,
\begin{align}
    P(x)=p_d P_d(x)+p_r P_r(x)\label{eq:fulldist0}
\end{align}
where 
\begin{align}
\label{eq:probweights}
p_r
&=\frac{r^{-1}}{\langle T \rangle_{x_0}+r^{-1}},
\quad p_d
=
\frac{\langle T \rangle_{x_0}}{\langle T \rangle_{x_0}+r^{-1}}.
\end{align}
This is compared to numerical simulations in Fig.~\ref{fig:fulldistnoninst} and shows very good agreement.

For intermediate values of $v_0$, the bath does not fully relax during the return, and the distribution of $q_0$ at restart retains broader memory of the pre-reset configuration. Increasing $v_0$ reduces the time available for bath relaxation and therefore broadens the effective distribution of $q_0$ sampled at the beginning of diffusive segments. This increases the typical magnitude of $z$ enhancing the subsequent spreading and yielding a broader stationary $P(x)$. 

\section{Conclusion}
\label{sec:conc}
In this work, we studied stochastic resetting of a Brownian probe embedded in a viscoelastic medium, focusing on regimes where the medium relaxation time is large and therefore directly influences the dynamical fluctuations. We introduced a minimal model for the viscoelastic medium in which it is represented by a single auxiliary bath particle coupled to the probe. Importantly, at each resetting event only the probe is reset, while the medium is left untouched. This is one of the main novelties of our work which is in sharp contrast to earlier works where the memory is also reset~\cite{biswas2025resetting} and the dynamics is renewed. The medium is of course affected indirectly because of the feedback coming from the probe, and this leads to a host of interesting results. 

For instantaneous resetting, we showed that in the regime of strong bath memory the stationary distribution is no longer exponential and instead develops Gaussian tails, which we characterized analytically. We obtained exact expressions for the time-dependent second moments and demonstrated a clear two-step relaxation toward stationarity in this regime, reflecting the separation between tracer–bath equilibration and slow bath relaxation.

We then considered non-instantaneous resetting via constant-velocity return. In contrast to Markovian Brownian motion, where the stationary distribution is invariant under the return protocol, we find that this invariance breaks down. The stationary distribution and variance depend sensitively on the return speed. Slow returns allow the bath to relax more before the next diffusive phase, leading to reduced fluctuations, while fast returns recover the instantaneous-reset limit. We identified analytically tractable limits of very large and very small return speeds, in which the stationary distributions become independent of the return velocity, and find that the typical relaxation time remains essentially unchanged. 
Our results reveal that, in the presence of environmental memory, the resetting protocol itself becomes a key control parameter governing the resulting non equilibrium steady states and can be tuned to design optimal search strategies in future. 

 Experimental realization in viscoelastic media appears feasible~\cite{ginot2026experimental}, particularly in setups where colloidal probes are driven at controlled velocities. The model presented here is minimal as it excludes the details of flow fields and specific relaxation mechanisms  but we expect the qualitative signatures predicted here to be observable in such systems. An obvious extension to our study will be including two or more bath particles which can mimic multiple relaxation~\cite{ginot2022recoil} and dissipation mechanisms~\cite{das2024friction} of the bath. It would also be interesting to explore baths with power-law memory kernels which naturally arise in spatially extended viscoelastic environments~\cite{jolakoski2025response}, disordered systems~\cite{das2022displacement} and in near critical fluid~\cite{demery2023non}. Future work involve investigating resetting in nonequilibrium baths where additional active fluctuations coupled with memory~\cite{santra2024forces} may produce even richer behavior. Finally, other forms of non-instantaneous resetting protocols may further elucidate the interplay between environmental memory and resetting.

\section{Acknowledgment}
IS acknowledges funding from the European Union’s Horizon
2024 research and innovation programme under the Marie Sklodowska-Curie (HORIZONTMA-MSCA-PF-EF) grant agreement No. 101205210. 
DD acknowleges the support by the Deutsche Forschungsgemeinschaft (DFG, German Research Foundation)—217133147/SFB 1073.

\appendix
\renewcommand{\theequation}{A\arabic{equation}}
\setcounter{equation}{0}
\renewcommand{\thefigure}{A\arabic{figure}}
\setcounter{figure}{0}
\section{Kurtosis}
Here, we present the expression of the kurtosis discussed in Sec~\ref{sec:instpos}. Starting from Eq.~\eqref{eq:momementgen}, the fourth order moments can be written as coupled differential equations similar to what was done for the second moments. 

Combining these equations with the second–order moment relations given in Eqs.~\eqref{eqs:2ndmom}, we obtain the expression  for the excess kurtosis $\kappa=\frac{M_{4,0}}{M_{2,0}}-3$ in the stationary state.
\begin{widetext}
\begin{align}
\label{eq:kurtosis}
   \kappa &=
   \frac {6 (\gamma r' + 
      1) (\gamma r' + 
       2)\left(\gamma\left(\gamma^2\left (r (r'+1)^2 + 
            3 \right) + \gamma (r (5 r' + 3) + 9) + 7 r' -5 \right) + 
      3 \right)} {(\gamma + \gamma r + 
        1)^2\left(\gamma\left (\gamma^2r' (r' + 1) (r' + 
            2) + \gamma (r ( r' + 19) + 36) + r'+1 \right) + 
      6 \right)} - 3
\end{align}
\end{widetext}
with $r'=r+2$. 

\begin{figure}
    \centering
    \includegraphics[width=0.69\linewidth]{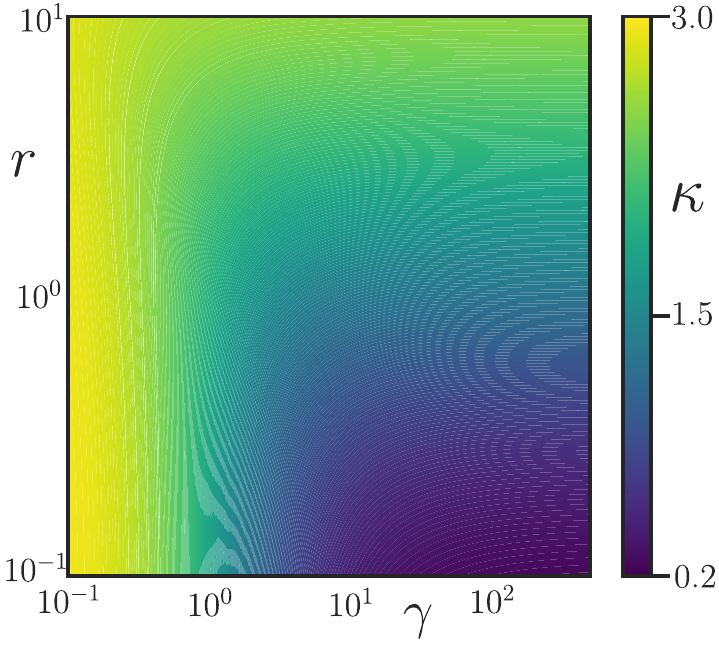}
    \caption{Density map of the kurtosis of the tracer position in the $(\gamma, r)$ parameter space. For small values of $\gamma$, the reset dominates and the kurtosis value is close to $3$ as expected from the resetting in Markovian fluid. For large $\gamma$ values, the value goes to $0$ }
    \label{fig:kurtosis}
\end{figure}
Fig. \ref{fig:kurtosis} presents the density map of $\kappa_4$ in the $(\gamma,r)$ plane, clearly revealing the reset-dominated and bath-dominated regimes.
\section{Steady state MSD in the limit $v_0\to 0$}
\renewcommand{\theequation}{B\arabic{equation}}
\setcounter{equation}{0}
Using Eq.~\eqref{eq:diffdist}, the stationary variance of the diffusing phase
\begin{align}
\nonumber
    \la x^2\ra_d&=r\int_0^\infty e^{-rt}\sigma_d^2(t) \\
    \nonumber
&=
\frac{1}{(\gamma+1)^2}
\left(
\frac{2\gamma^2\lambda}{r+2\lambda}
+\frac{2(\gamma+1)}{r}
\right)\\
&+
\frac{1}{\lambda^2}
\left(
1-\frac{2r}{r+\lambda}
+\frac{r}{r+2\lambda}
\right)
\end{align}
To obtain the stationary mean-square displacement (MSD) during the return phase, we use the position distribution, $P_r(x)=\frac{1}{\langle |x_0|\rangle}\int_{|x'|>|x|} P_d(x') dx$. The MSD in the return phase can then be written as
\begin{align}
\nonumber
    \langle x^2\rangle_r&=
\frac{1}{\langle |x_0|\rangle}
\int_{-\infty}^{\infty}x^2
\left[
\int_{|x'|>|x|}P_d(x') dx'
\right]dx \\
\nonumber
&= \frac{1}{\langle |x_0|\rangle}
\int_{-\infty}^{\infty}
P_d(x')\frac{2|x'|^3}{3} dx \\
&= \frac{\langle |x_0| \rangle^3}{3 \langle |x_0| \rangle}
\end{align}
Substituting the expression for $P_d(x)$ gives
\begin{align}
\langle x^2 \rangle_r\ = \frac{2}{3}
\frac{\int_{0}^{\infty} r e^{-rt}\,\sigma_d^{3}(t)\,dt}
{\int_{0}^{\infty} r e^{-rt}\,\sigma_d(t)\,dt}
\end{align}
The total steady state MSD can then be calculated using Eq.~\eqref{eq:fulldist0},
\begin{align}
    \langle x^2\rangle_{ss} = p_r \langle x^2 \rangle_r + p_d \langle x^2 \rangle_d\label{eq:noninst:msd}
\end{align}
The stationary value of MSD for $r = 0.1$ is shown in Fig~\ref{fig:msd_v}({\it Inset}).

\bibliography{reset}

\end{document}